\documentclass[a4paper]{article}
\usepackage{epsfig}
\usepackage{graphicx}  
\usepackage{color}     
\usepackage{lscape}
\usepackage{color}
\date{\today}
\def\a{\alpha}
\def\b{\beta}

\def\n{\nu}
\def\m{\mu}
\def\d{\delta}
\def\D{\Delta}
\def\s{\sigma}
\def\r{\rho}
\def\t{\theta}

\def\la{\lambda}

\def\arnp#1#2#3{     {\it Ann. Rev. Nucl. Part. Sci.}{\bf{#1},} #2 (#3)  }

\def\ijmpa#1#2#3{  {\it Int. J. Mod. Phys.} {A \bf{#1},} #2 (#3)  }

\def\plb#1#2#3{    {\it Phys. Lett. }{ B \bf{#1},}#2 (#3)  }

\def\prd#1#2#3{    {\it Phys. Rev. }{ D \bf{#1},} #2 (#3)  }

\def\prl#1#2#3{    {\it Phys. Rev. Lett. }{\bf{#1},} #2 (#3)  }
\def\mpla#1#2#3{    {\it Mod. Phys. Lett. }{A \bf{#1},} #2 (#3)  }

\def\ppnp#1#2#3{    {\it Prog. Part. Nucl. Phys. }{\bf{#1},} #2 (#3)  }

\def\ibid#1#2#3{   {\it ibid. }{\bf{#1},} #2 (#3)  }

\def\eq#1{{equation~(\ref{#1})}}

\def\etal{{\it et al.}}

\newcommand{\bea}{\begin{eqnarray}}
\newcommand{\beq}{\begin{equation}}
\newcommand{\eea}{\end{eqnarray}}
\newcommand{\eeq}{\end{equation}}
\newcommand{\nnu}{\nonumber}
\begin{document}
\title{\bf Textures with two traceless submatrices of the neutrino mass matrix }
{\small
\author{H. A. Alhendi$^1$\ , \ E. I. Lashin$^{1,2,3}$ and \ A. A. Mudlej$^1$\\
$^1$ Department of physics and Astronomy, College of Science,\\ King Saud University, Riyadh,
Saudi Arabia \\
$^2$ The Abdus Salam ICTP, P.O. Box 586, 34100 Trieste, Italy\\
$^3$ Department of Physics, Faculty of Science, \\Ain Shams University, Cairo, Egypt\\
Emails: alhendi@ksu.edu.sa, elashin@ictp.it and lashin@ksu.edu.sa}
}
\maketitle
\begin{abstract}
We propose a new texture for the light neutrino mass matrix. The
proposal is based upon imposing zero-trace condition on the two by
two sub-matrices of the complex symmetric Majorana mass matrix
in the flavor basis where the charged lepton mass matrix is diagonal.
Restricting the mass matrix to have two traceless sub-matrices may
be found  sufficient to describe the current data. Eight
out of fifteen independent possible cases are  found to be compatible
with current data. Numerical and some approximate
analytical results are presented.
\end{abstract}

\section{Introduction}
The observed phenomenon of neutrino oscillations \cite{sk}--\cite{chooz}, solar
and atmospheric,
provides a compelling  evidence that neutrinos are massive and lepton flavors
are mixed. These facts are in contrast to the Standard Model of Particle Physics,
especially the Electro-weak interaction which is based on $SU(2)_L\times U(1)_Y$ gauge,
where the neutrinos are massless. We will assume that the
neutrinos are of Majorana type neutrinos, as favored by some theoretical
considerations\cite{kayser}, whence the mass matrix $M$ is symmetric.
In the frame work of
three lepton families, the mass spectrum and flavor mixing are fully described
by twelve real parameters: three charged lepton masses $(m_e, m_\mu, m_\tau)$,
three neutrino masses $(m_1, m_2, m_3)$, three flavor mixing angles
$(\t_x, \t_y, \t_z)$, one Dirac-type CP-violating phase $(\d)$ and two
Majorana-type CP-violating phases $(\r\; \mbox{and}\; \s)$.

In the flavor basis where the charged lepton mass matrix is diagonal,
the mass term for Majorana neutrinos in terms of gauge eigen states,
in the case of three flavors, has the form
\beq
{\cal L}_{\mbox{mass}} = -{1\over 2}\,\n_L^T\,C^{-1}\,M^\dagger\,\n_L + h.c,
\label{nmass}
\eeq
where $C$ is the charge conjugation matrix. The complex symmetric Majorana
neutrino mass matrix $M$ can be diagonalized by unitary
transformation that links the gauge and mass eigen states as:
\beq
\left(
\begin{array}{l}
\n_e \\
\n_\m \\
\n_\tau
\end{array}
\right)
=
\left(
\begin{array}{lll}
V_{e1} & V_{e2} & V_{e3}\\
V_{\m 1} & V_{\m 2} & V_{\m 3}\\
V_{\tau 1} & V_{\tau 2} & V_{\tau 3}
\end{array}
\right)
\left(
\begin{array}{l}
\n_1' \\
\n_2' \\
\n_3'
\end{array}
\right).
\label{mix0}
\eeq
The mass term can be written in terms of mass eigen states as
\beq
{\cal L}_{\mbox{mass}} = -{1\over 2}\,{\n'}_L^T\,C^{-1}\,D\,
{\n'}_L + h.c,
\label{n1mass}
\eeq
where
\beq
V^{T}\,M^\dagger\,V= D
\label{mv}
\eeq
and $D=\mbox{diag}(m_1,m_2,m_3)$ with $m_i$  real positive numbers.
The lepton flavor mixing matrix $V$ contains six real parameters,
three of them are mixing angles while the rest are three
CP-violating phases. Following the parameterization in \cite{xing1},
the matrix $V$ can be expressed as a product
of the Dirac-type flavor mixing matrix $U$ (consisting of three
mixing angles and one CP-violating phase) and a diagonal matrix $P$
(consisting of two nontrivial Majorana phases): $V=U\,P$. Then we
may rewrite $M$ in \eq{mv} as
\beq
M = U
\left(
\begin{array}{lll}
\la_1 & 0 & 0\\
0 & \la_2 & 0\\
0 & 0 & \la_3
\end{array}
\right)
U^T
\label{mass}
\eeq
where two Majorana-type CP-violating phases are included into the
complex neutrino mass eigenvalues $\la_i$, and the relation
$\left|\la_i\right|=m_i$ holds. Without loss of generality, we take
\beq
\la_1 = m_1\,e^{2\,i\,\rho}\;\; ,\;\; \la_2 =
m_2\,e^{2\,i\,\sigma}\;\; , \;\; \la_3 = m_3
\label{la}
\eeq

The matrix $U$ is the Dirac-type flavor mixing matrix \cite{xing1}
\beq
U=
\left(
\begin{array}{ccc}
c_{x}\,c_{z}  & s_{x}\,c_{z} & s_{z} \\
-c_{x}\,s_{y}\,s_{z} - s_{x}\,c_{y}\,e^{-i\,\d}& s_{x}\,s_{y}\,s_{z}
+c_{x}\,c_{y}\,e^{-i\,\d} &s_{y}\,c_{z} \\
-c_{x}\,c_{y}\,s_{z} + s_{x}\,s_{y}\,e^{-i\,\d} &
 -s_{x}\,c_{y}\,s_{z} - c_{x}\,s_{y}\,e^{-i\,\d} & c_{y}\,c_{z} \\
\end{array}
\right)
\label{vp}
\eeq
where we used the convention $s_{\a}=\sin{\a}$ , $c_{\a}=\cos{\a}$.

For the parameterization followed in the present work,
the mixing angles $(\t_x, \t_y, \t_z)$
are directly related to the angles  of solar, atmospheric and
CHOOZ reactor oscillations\cite{xing1}:
\beq
\t_x\approx \t_{\mbox{sol}},\;\;\t_y\approx
\t_{\mbox{atm}},\;\; \t_z\approx \t_{\mbox{chz}},
\label{xyz}
\eeq
and
\beq
\D m_{\mbox{sol}}^2 = \left|m_2^2 - m_1^2\right|,\;\;
\D m_{\mbox{atm}}^2 = \left|m_3^2 - m_1^2\right|.
\label{dm1}
\eeq

Beta decay, neutrinoless double-beta decays and precision cosmology are sensitive
to the absolute neutrino mass scale. The dependence can be respectively characterized
through two non-oscillation parameters and mass sum  parameter as follows:
the effective electron neutrino mass term
\beq
M_\b = \sqrt{\left|U_{e1}\right|^2 m_1^2 + \left|U_{e2}\right|^2 m_2^2
+ \left|U_{e3}\right|^2 m_3^2},
\label{effme}
\eeq
the effective mass term of neutrinoless double beta decay
\beq
M_{\b\b} =
m_3\,\left|\frac{m_1}{m_3}\,U_{e1}^2\,e^{2\,i\,\rho} +
 \frac{m_2}{m_3}\,U_{e2}^2\,e^{2\,i\,\sigma}
+ U_{e3}^2\right|,
\label{nless}
\eeq
and the sum mass parameter
\beq
\Sigma = m_1 + m_2 + m_3.
\label{msum}
\eeq

A recent global analysis of neutrino oscillation data\cite{fog}, at the confidence level
of $95\%$, gives the best estimates of the oscillation parameters as
\beq
\begin{array}{lll}
\D m_{\mbox{atm}}^2 &=& (2.4^{+0.5}_{-0.6})\times 10^{-3}
\mbox{eV}^2,\\
\D m_{\mbox{sol}}^2  &=& (7.92\pm 0.7)\times 10^{-5} \mbox{eV}^2,
\\
\sin^2{\t_{\mbox{sol}}} &=&0.314^{+0.057}_{-0.047},\\
\sin^2{\t_{\mbox{atm}}} &=&0.44^{+0.18}_{-0.01},\\
\sin^2{\t_{\mbox{chz}}}&=&0.9^{+2.3}_{-0.9}\times 10^{-2},
\end{array}
\label{osp}
\eeq
where $\t_{\mbox{atm}}$ and $\t_{\mbox{sol}}$ are the angles relevant
for atmospheric  and solar neutrino oscillation  respectively.
While, $\t_{\mbox{chz}}$ is the angle relevant to CHOOZ
reactor experiment for neutrino oscillation\cite{chooz}. A useful
parameter $R_\n$ can be defined as:
\beq
R_\n ={\D m_{\mbox{sol}}^2 \over \D m_{\mbox{atm}}^2},
\label{defR}
\eeq
which has possible values ranging, at the level of confidence $95\%$,
as
\beq
0.025 \le R_\n\le 0.049,
\label{cons}
\eeq
this constraint constitutes a very tight criteria for the model to
be phenomologically acceptable.

Whereas the three masses of charged leptons $(m_e, m_\m, m_\n)$ have
precisely been measured\cite{pdg}, we have, concerning the absolute neutrino mass
scale, only experimental bounds for $M_\b$,  $M_{\b\b}$  and $\Sigma$ as follow (see
\cite{fog} and refs. therein)
\beq
\begin{array}{ll}
\left.
\begin{array}{lll}
M_\b &<&  1.8 \;\mbox{eV}\\
M_{\b\b} &=& 0.58^{+0.22}_{-0.16}\; \mbox{eV}\\
\Sigma &<& 1.4\; \mbox{eV}
\end{array}
\right\}
&95\% \;\mbox{confidence level}.
\end{array}
\label{neq}
\eeq
The lower bound for $M_{\b\b}$ disappears in case of absence of the neutrinoless
double-beta decay. At present, there is no available precise information on any of the
CP violating phases.

The present available data on neutrinos, even those in the foreseeable future,
can not fully determine all the parameters in the neutrino mass matrix.
A challenging theoretical task is to find out a Majorana mass matrix of
the light neutrino consistent with the current data as summarized in \eq{osp}. Several
attempts have been made to obtain phenomenologicaly acceptable
patterns of the neutrino mass matrix M, such as texture zeros
\cite{fram1}--\cite{xing2}, Zero sum condition~\cite{zee},and determinant zero
requirement \cite{branco} for the mass matrix. There are also many proposal for
the mass matrix based on some symmetry group as in \cite{ma}.

In this work we impose the condition that the trace of two possible
$2\times 2$ submatices is zero. The traceless condition can be considered
as a non trivial generalization of the zero-textures, since that,
a zero-element can be viewed as a zero-trace of a $1 \times 1$ sub-matrix.
Taking these submatrices in pairs we
obtain $15$ independent possibilities. Out of these, we find
just $8$ of them to be phenomenologicaly acceptable. However there are other
five cases which can be considered to be marginally accepted. The
numerical and some analytical approximate results are reported for
all these thirteen cases.

The plan of the paper is as follows: in section $2$, necessary formulas for
the calculations are introduced beside a classification of the resulting
mass patterns.
Sections~$3$, $4$ and~$5$ are respectively devoted to the resulting three possible
mass patterns. For each pattern, we present the textures of $M$
with two independent vanishing traces and compute the expressions
of the two neutrino mass ratios and the Majaorana phases and other relevant parameters.
Numerical and some approximate analytical results are
presented. Consistency of models with experimental results are
discussed. We end up by conclusions and discussions in section~5.

\section{Fifteen Possible texture with two Traceless submatrices}
As $M$ is $3\times 3$ symmetric complex matrix, it totally has six independent
complex entries. If we impose the condition that the trace of two possible
$2\times 2$ submatrices is zero, then we have in total $6$ independent
submatrices with zero trace. When these independent submatrices
are taken into pairs, we get $15$ possibilities that can be
written as
\bea
M_{rs} + M_{ij}&=&0 \nnu\\
M_{\a\b} + M_{nm} &=& 0.
\eea
where each subscript runs over $e (1)$, $\mu (2)$ and $\tau (3)$, but $(r s)\neq
(ij)$ and $(\a\b)\neq (nm)$.

Using \eq{mass} we then obtain the following constraint relations:
\bea
\sum_{l=1}^{3}\left( U_{r l}\,U_{s l} +
U_{il}\,U_{jl}\right)\lambda_l &=& 0 \nnu \\
\sum_{l=1}^{3}\left( U_{\a l}\,U_{\b l} +
U_{nl}\,U_{ml}\right)\lambda_l &=& 0
\label{cond1}
\eea
The solutions of \eq{cond1} can be written as:
\bea
\frac{\la_1}{\la_3} &=& \frac{a_3\,b_2 -a_2\, b_3}{b_1\,a_2 -
a_1\,b_2} \nnu \\
\frac{\la_2}{\la_3} &=& \frac{a_1\,b_3 -a_3\, b_1}{b_1\,a_2 -
a_1\,b_2}
\label{sol1}
\eea
where
\bea
a_l &=& U_{rl}\,U_{sl} + U_{il}\,U_{jl} \nnu \\
b_l &=& U_{\a l}\, U_{\b l} + U_{nl}\,U_{ml}.
\label{abs}
\eea
One can observe that the left-hand sides of \eq{sol1}
contain Majorana-type CP-violating phases, while the
right-hand sides contain the Dirac-type
CP-violating phase. Therefore two
Majorana phases must depend upon the Dirac-type CP-violating
phase. This dependence results simply from the texture of tracelss
submatrices that we have taken.

\begin{figure}[hbtp]
\centering
\epsfxsize=6cm
\centerline{\epsfbox{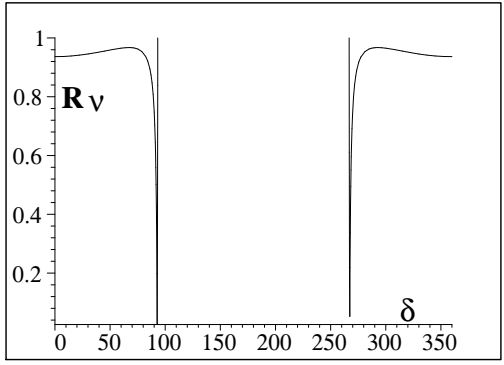}}
\caption{{\footnotesize  $R_\n$ as a function of $\d$, whereas $\t_x=34^0$,
 $\t_y=42^0$ and $\t_z=5^0$ for the model ${\bf D_1}$. For illustrative purposes the vertical range is restricted
 to the interval from zero to one. The sharp singling out of $\d=93^0$  and $\d=268^0$
 as consistent values with the relation $0.025 \le R_\n \le 0.049$ is clear.}}
\label{selc}
\end{figure}

Comparing \eq{sol1} with \eq{la}, we get the two neutrino mass
ratios :
\bea
\frac{m_1}{m_3} &=& \left|\frac{a_3\,b_2 -a_2\, b_3}{b_1\,a_2 -
a_1\,b_2}\right|, \nnu \\
\frac{m_2}{m_3} &=&\left| \frac{a_1\,b_3 -a_3\, b_1}{b_1\,a_2 -
a_1\,b_2}\right|,
\label{solm}
\eea
and the two Majorana phases:
\bea
\rho &=& \frac{1}{2}\;\mbox{arg}\left[\frac{a_3\,b_2 -a_2\, b_3}{b_1\,a_2 -
a_1\,b_2}\right], \nnu \\
\sigma &=&\frac{1}{2}\;\mbox{arg}\left[ \frac{a_1\,b_3 -a_3\, b_1}{b_1\,a_2 -
a_1\,b_2}\right].
\label{solp}
\eea
With  the inputs of three flavor mixing angles and the Dirac-type
CP-violating phase, we would be able to predict the relative magnitude
of three neutrino masses, the values of two Majorana phases and $R_\n$.
The absolute neutrino mass scale can then be predicted by matching,
for example, the value of $\D m^2_{\mbox{sol}}$. The other remaining
parameters $M_\beta$, $M_{\beta\beta}$ and $\Sigma$ can also be predicted.
This predictability allows us to examine whether the chosen
texture of $M$ with two traceless submatrices is empirically
acceptable or not. The input values of $\t_x$ and $\t_y$ should be
consistent with the bounds given by \eq{osp}, typical ones can
be taken as $\t_x = 34^0$ and $\t_y=42^0$. The Dirac-type
CP-violating phase is not constrained. The strategy followed to
obtain a good choice for $\d$ is to plot the parameter $R_\n$ as
a function of $\d$ while maintaining $\t_x\approx 34^0$,
$\t_y=42^0$ and $\t_z\approx 5^0$. The constraint in \eq{cons}
turns out to be generically very selective for the appropriate choice of $\d$,
as shown in Fig.~\ref{selc} for a particular model $\bf D_1$ (according to the
nomenclature explained later).

The resulting mass patterns turn out to be classified according
to the following three classes:
\begin{itemize}
\item Degenerate case which is characterized by $m_1\sim m_2 \sim m_3$
and is denoted by ${\bf D}$.
\item Normal hierarchy which is characterized by $m_1\sim m_2 < m_3$
and is denoted by ${\bf N}$.
\item Inverted hierarchy which is characterized by $m_1\sim m_2 > m_3$
and is denoted by ${\bf I}$.
\end{itemize}
In all our subsequent discussion we follow this order and nomenclature.

To work out the explicit expressions of
${\la_1/ \la_3}$ and ${\la_2/ \la_3}$ in each case, we adopt
the parameterization  given in \eq{vp} for the Dirac-type flavor
mixing matrix, from which we then obtain
the analytical results
for $m_1/m_3,\; m_2/ m_3,\; \rho ,\; \sigma,\; R_\nu,\;M_\beta$ and
$M_{\beta\beta}$. However the analytical results as well as
the approximate one (for small $s_z$) are too lengthy to be
displayed here, and thus we quote only the numerical results based
on the exact formulae in \eq{sol1}. We present, for each model,
the expressions of $a$'s and $b$'s coefficients and the analytic approximate
results for $R_\n$. However for the sake of presentation,
the approximate analytical results of all parameters are presented for
just one case for each pattern class.

\section{Degenerate models}
\begin{figure}[hbtp]
\centering
\begin{minipage}[c]{0.5\textwidth}
\epsfxsize=6cm
\centerline{\epsfbox{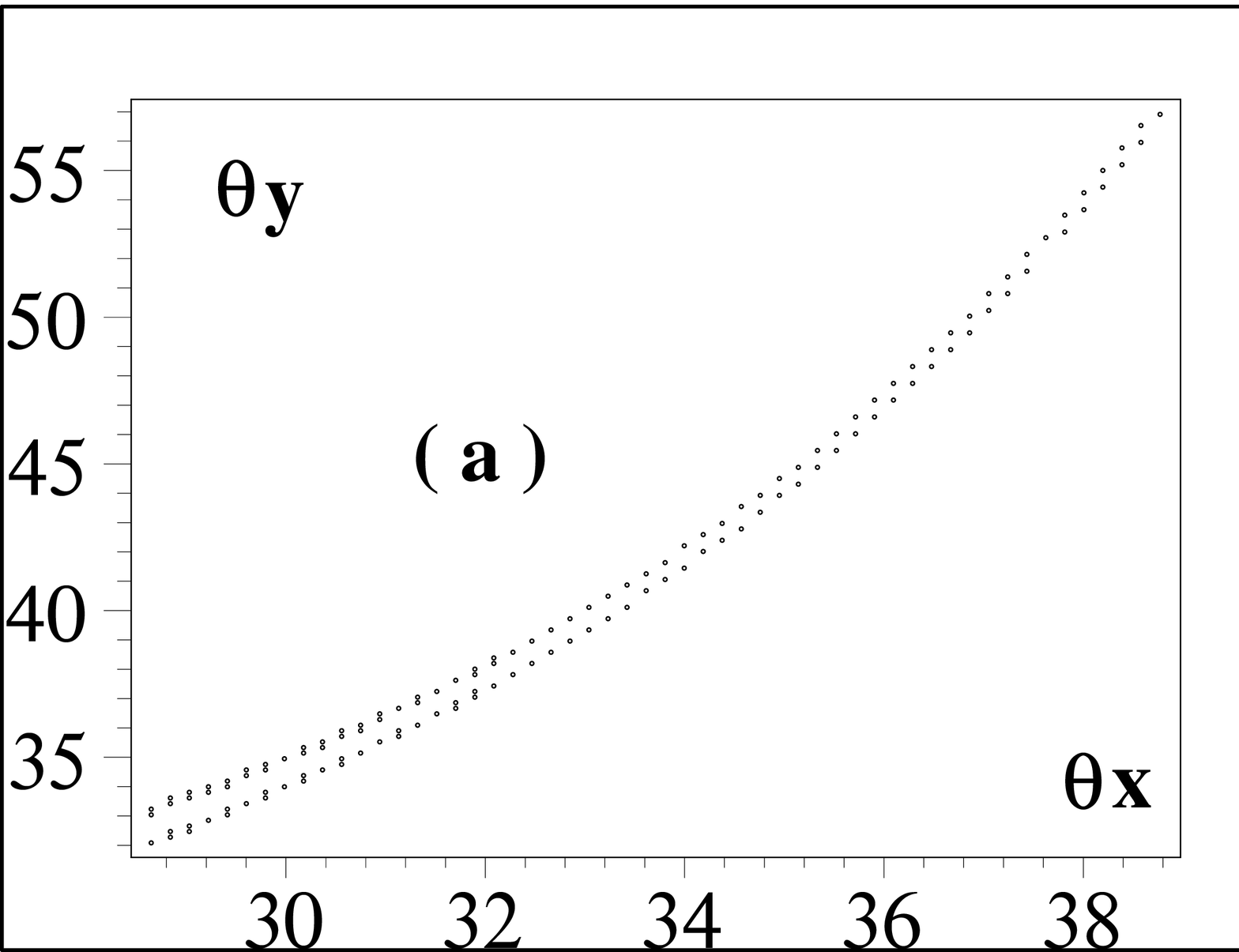}}
\end{minipage}%
\begin{minipage}[c]{0.5\textwidth}
\epsfxsize=6cm
\centerline{\epsfbox{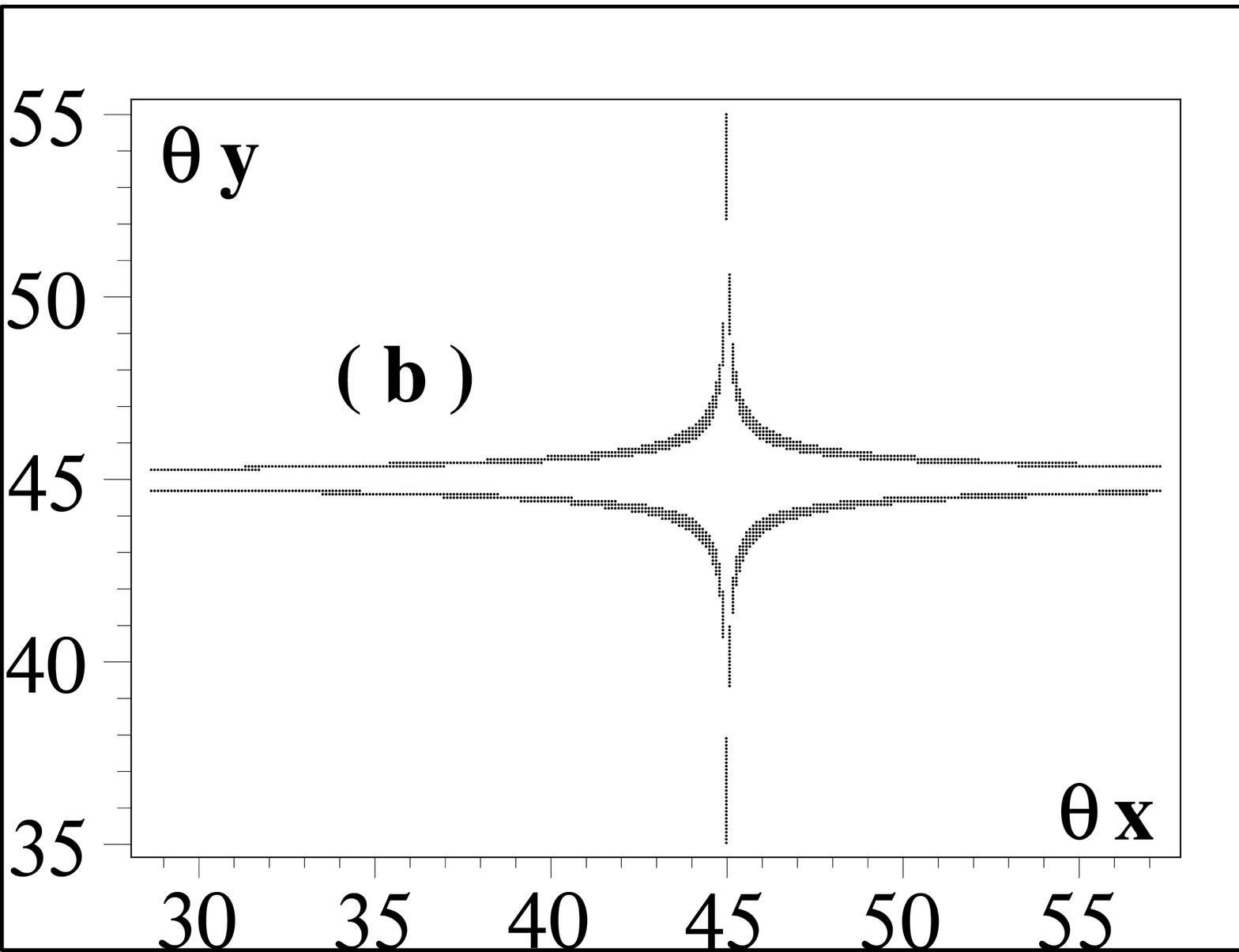}}
\end{minipage}
\caption{{\footnotesize  The available parameter space for the cases
$\bf D_1$ (a)  with $\d=92.7^0$ and $\bf D_2$ with $\d=90^0$ (b) respectively.
The dotted region represents the  points $(\t_x, \t_y)$ at
which $R_\n$ is constrained by $0.025 < R_\n <0.05$. The range from $29^0$ to
$57^0$ is spanned for both angles $\t_x$ and $\t_y$.  }}
\label{scfig}
\end{figure}

{\bf Pattern} $\bf D_1$: $M_{ee} + M_{\mu\mu} = 0, M_{e\mu} + M_{\mu\tau}
=0$. In this pattern the required quantities $a$'s and $b$'s as given
by \eq{abs} are
\bea
&& a_1=c_x^2 c_z^2+(-c_x s_y s_z-s_x c_y e^{-i\,\d})^2,\;\;
 a_2=s_x^2 c_z^2+(-s_x s_y s_z+c_x c_y e^{-i\,\d})^2,\;\;
a_3=s_z^2+s_y^2 c_z^2, \nnu \\
&& b_1=c_x c_z (-c_x s_y s_z-s_x c_y e^{-i\,\d})+(-c_x s_y s_z-s_x c_y e^{-i\,\d})
(-c_x c_y s_z+s_x s_y e^{-i\,\d}),\nnu \\
&& b_2=s_x c_z (-s_x s_y s_z+c_x c_y e^{-i\,\d})+(-s_x s_y s_z+c_x c_y e^{-i\,\d})
(-s_x c_y s_z-c_x s_y e^{-i\,\d}),\nnu \\
&& b_3=s_z s_y c_z+s_y c_z^2 c_y ,
\label{abd1}
\eea
which are sufficient to calculate all other quantities.
The corresponding expression for  $R_\n$, expanded at
the leading power of $s_z$, is
\beq
R_\n\approx \left|{-s_x s_y^3\over
2 c_x c_\d  s_y^2 s_x^2-c_x c_\d s_x^2+s_x^3 s_y^3-2 s_x^3 s_y+s_x s_y}\right|
+O(s_z).
\eeq
In this pattern, to match the experimental results, the required
values are $(\t_x = 34^0, \t_y = 42^0, \d = 92.755^0, \t_z = 5^0)$.
For these inputs we obtain $m_1/m_3 = 1.050052527$, $m_2/m_3 =
1.048504453$, $\r  = 87.72^0$, $\s = 95^0$ and $R_\nu = 0.033$.
The mass $m_3$ fitted from the observed $\D m^2_{\mbox{sol}}$ is $m_3 = 0.156\;
\mbox{eV}$. Then the derived values for the other remaining
parameters are $\D m^2_{\mbox{atm}}=2.5\times 10^{-3}\;\mbox{eV}^2$, $M_\b
=0.164\;\mbox{eV}$, $M_{\b\b} =0.160 \;\mbox{eV}$ and $\Sigma =0.48 \;\mbox{eV} $.
There is no tuning required for the mixing angles $\t_x$ and $\t_y$,
to assure their consistency with the relation $0.025 < R_\n <0.05$
as is shown in Fig.~\ref{scfig}(a). In this pattern the numerically
estimated mass matrix $M$ is
\beq
M=
m_3\left(
\begin {array}{ccc} - 1.0268+ 0.000489\,i& 0.027333- 0.00049\,i& 0.21400+ 0.00037\,i
\\\noalign{\medskip}
0.027333- 0.00049\,i&1.0268- 0.000513\,i&- 0.02737+ 0.000517\,i
\\\noalign{\medskip}
 0.21400+ 0.00037\,i&- 0.02737+ 0.000517\,i& 0.99942- 0.000507\,i
\end {array}
\right)
\label{massa1}
\eeq

{\bf Pattern} $\bf D_2$: $M_{ee} + M_{\tau\tau} = 0, M_{ee} + M_{\mu\mu} =0.$
In this pattern the required quantities $a$'s and $b$'s as given
by \eq{abs} are
\bea
a_1=c_x^2 c_z^2+(-c_x c_y s_z+s_x s_y e^{-i\,\d})^2,
& a_2=s_x^2 c_z^2+(-s_x c_y s_z-c_x s_y e^{-i\,\d})^2,
& a_3=s_z^2 + c_z^2 c_y^2,\nnu\\
b_1=c_x^2 c_z^2+(-c_x s_y s_z-s_x c_y e^{-i\,\d})^2,
& b_2=s_x^2 c_z^2+(-s_x s_y s_z+c_x c_y e^{-i\,\d})^2,
& b_3=s_z^2+s_y^2 c_z^2,
\label{abn3}
\eea
which are sufficient to calculate all other quantities.

Using $s_z$ as a small parameter, we expand in terms of its powers and keep
only leading terms.
The analytical approximate formula for the mass ratio are:
\beq
{m_1\over m_3} \approx {m_2\over m_3}\approx c_{2x}^{-1}\;\sqrt{1-4 s_x^2 c_x^2 s_\d^2 } + O(s_z)\nnu \\
\eeq

The corresponding expressions for $\r$ and $\s$ are
\bea
\r \approx -{1\over 2} \tan^{-1}\left({2 s_x^2 c_\d s_\d \over
-1+2 s_x^2 s_\d^2}\right) +
O(s_z), & &
\s \approx {1\over 2} \tan^{-1}\left({2 c_x^2 c_\d s_\d \over
1-2 c_x^2 s_\d^2}\right) +
O(s_z),
\eea
while the corresponding expressions for $R_\n$, $M_{\b\b}$ and $M_\b$
are
\beq
R_\n\approx \left|{2 c_y s_y s_x\over
c_\d c_x (1+2 c_x^2 c_y^2-2 c_y^2-c_x^2)}\right| s_z +O(s_z^2)
\eeq
and
\bea
M_{\b\b}=m_3+ O(s_z), & & M_{\b}=m_3 \;\sqrt{1-4 s_x^2 s_\d^2 c_x^2
\over 1-4 s_x^2 c_x^2} + O(s_z).
\eea

In this pattern, to match the experimental results, the required
values are $(\t_x = 34^0, \t_y = 44.65^0, \d = 90^0, \t_z = 5^0)$.
For these inputs we obtain $m_1/m_3 = 1.015274786$, $m_2/m_3 =
1.014764820$, $\r  = 89.99^0$, $\s = 89.86^0$ and $R_\nu = 0.035$.
The mass $m_3$ fitted from the observed $\D m^2_{\mbox{sol}}$ is $m_3 = 0.28\;
\mbox{eV}$. Then the derived values for the other remaining
parameters are $\D m^2_{\mbox{atm}}=2.4\times 10^{-3}\;\mbox{eV}^2$, $M_\b
=0.28\;\mbox{eV}$, $M_{\b\b} =0.28 \;\mbox{eV}$ and $\Sigma =0.84 \;\mbox{eV} $.
The angle $\t_y$  is highly constrianed around $45^0$ for a
reasonable choice of $\t_x$ as can be seen from
Fig.~\ref{scfig}~(b).
In this pattern the numerically
estimated mass matrix $M$ is

\beq
M=
m_3\,\left(
\begin {array}{ccc}
- 0.99982+ 0.0017117\,i& 0.12446- 0.00027\,i& 0.12299+ 0.00004\,i\\
 0.12446- 0.00027\,i& 0.99981- 0.001714\,i&- 0.01515+ 0.0017276\,i\\
 0.12299+ 0.00004\,i&- 0.01515+ 0.0017276\,i& 0.99982- 0.001714\,i
\end {array}
\right)
\label{massa4}
\eeq

{\bf Pattern} $\bf D_3$: $M_{\mu\mu} + M_{\tau\tau} = 0, M_{\mu e} +
M_{\tau\tau} =0.$ In this pattern the required quantities $a$'s and $b$'s as given
by \eq{abs} are
\bea
&& a_1=c_x^2 s_z^2+s_x^2 e^{-2\,i\,\d},\;\;
 a_2=s_x^2 e^{-2\,i\,\d} +s_x^2 s_z^2,\;\;
a_3=c_z^2, \nnu \\
&& b_1=(-c_x s_y s_z-s_x c_y e^{-i\,\d}) c_x c_z+(-c_x c_y s_z
+s_x s_y e^{-i\,\d})^2,\nnu \\
&& b_2=(-s_x s_y s_z+c_x c_y e^{-i\,\d}) s_x c_z
+(-s_x c_y s_z-c_x s_y e^{-i\,\d})^2,\nnu \\
&& b_3=s_y s_z c_z+c_y^2 c_z^2 ,
\label{abd3}
\eea
which are sufficient to calculate all other quantities.
The corresponding expression for  $R_\n$, expanded at
the leading power of $s_z$, is
\beq
R_\n\approx \left|{2 c_\d  c_x c_y s_x-4 s_y^2 s_x^2+2 s_x^2+2 s_y^2-1\over
-2 c_\d c_x c_y s_x^3+2 s_y^2 s_x^4-s_x^4}\right|
+O(s_z).
\eeq
In this pattern, to match the experimental results, the required
values are $(\t_x = 34^0, \t_y = 42^0, \d = 272.9^0, \t_z = 5^0)$.
For these inputs we obtain $m_1/m_3 = 1.008141341$, $m_2/m_3 =
1.007871889$, $\r  = 177.55^0$, $\s = 5.32^0$ and $R_\nu = 0.034$.
The mass $m_3$ fitted from the observed $\D m^2_{\mbox{sol}}$ is $m_3 = 0.38\;
\mbox{eV}$. Then the derived values for the other remaining
parameters are $\D m^2_{\mbox{atm}}=2.4\times 10^{-3}\;\mbox{eV}^2$, $M_\b
=0.38\;\mbox{eV}$, $M_{\b\b} =0.38\;\mbox{eV}$ and $\Sigma =1.15 \;\mbox{eV} $.
In this pattern the numerically
estimated mass matrix $M$ is
\beq
M=
m_3\,\left(
\begin {array}{ccc}  1.0001- 0.000939\,i&- 0.093575- 0.00001\,i& 0.084241+ 0.00011\,i\\\noalign{\medskip}- 0.093575- 0.00001\,i&-
 0.09358+ 0.000006\,i& 0.99567- 0.000001\,i\\\noalign{\medskip}
 0.084241+ 0.00011\,i& 0.99567- 0.000001\,i& 0.09355- 0.000011\,i
\end {array}
\right)
\label{massa9}
\eeq

\section{Normal hierarchy models}
{\bf Pattern} $\bf N_1$: $M_{e\mu} + M_{\mu\tau} = 0, M_{ee} + M_{\mu\tau} =0.$
In this pattern the required quantities $a$'s and $b$'s as given
by \eq{abs} are
\bea
a_1&=&c_x c_z (-c_x s_y s_z-s_x c_y e^{-i\,\d})+
(-c_x s_y s_z-s_x c_y e^{-i\,\d}) (-c_x c_y s_z+s_x s_y e^{-i\,\d})\nnu\\
a_2&=&s_x c_z (-s_x s_y s_z+c_x c_y e^{-i\,\d})+
(-s_x s_y s_z+c_x c_y e^{-i\,\d}) (-s_x c_y s_z-c_x s_y e^{-i\,\d})\nnu\\
a_3&=&s_z s_y c_z+s_y c_z^2 c_y\nnu\\
b_1&=&c_x^2 c_z^2+
(-c_x s_y s_z-s_x c_y e^{-i\,\d}) (-c_x c_y s_z+s_x s_y e^{-i\,\d})\nnu\\
b_2&=&s_x^2 c_z^2+
(-s_x s_y s_z+c_x c_y e^{-i\,\d}) (-s_x c_y s_z-c_x s_y e^{-i\,\d})\nnu\\
b_3&=&s_z^2+s_y c_z^2 c_y,
\label{abn1}
\eea
which are sufficient to calculate all other quantities.

Using $s_z$ as a small parameter, we expand in terms of its powers and keep
only the leading terms.
The analytical approximate formulae for the mass ratios are:
\bea
{m_1\over m_3} \approx s_y s_x\;\sqrt{ 2 s_x c_x c_\d c_y-c_y^2-s_x^2 s_y^2\over N_m} + O(s_z),
&&{m_2\over m_3} \approx c_x\, s_y\;\sqrt{ -2 c_\d c_y c_x s_x+s_x^2 s_y^2-1\over N_m} + O(s_z)
\eea
where
\bea
N_m&=&4 c_\d s_x^3 c_x c_y s_y^2-2 s_y^2 c_y s_x c_x c_\d-4 c_\d s_y s_x^3 c_x
+2 c_\d s_y s_x c_x+4 c_y s_y s_\d^2 s_x^4
-2 c_y s_y s_x^4-4 c_y s_y s_\d^2 s_x^2\nnu \\
&&+2 c_y s_y s_x^2-s_y^4 s_x^4-3 s_x^4 s_y^2+s_x^4+s_y^4 s_x^2+3 s_x^2 s_y^2-s_x^2-s_y^2,
\eea

The corresponding expression for $\r$ and $\s$ are
\bea
\r \approx  {1\over 2} \tan^{-1}\left({N_{\r1}\over N_{\r2}}\right) +
O(s_z), &&
\s \approx {1\over 2} \tan^{-1}\left({N_{\s1}\over N_{\s2}}\right) +
O(s_z),
\eea
where
\bea
N_{\r1}&=&s_x s_\d (-4 c_\d^2 c_y c_x s_y+c_x^3 c_y s_y+4 c_x^3 c_y c_\d^2 s_y
+c_x-c_x^3+2 c_\d s_y s_x-4 c_\d s_y s_x c_x^2+2 c_\d s_y s_x c_x^2 c_y^2),\nnu\\
N_{\r2}&=& (4 s_\d^2 c_\d-3 c_\d) c_x s_y c_y s_x^3+c_y s_x^4+
(1-2 s_\d^2) (s_y s_x^4- s_y^3 s_x^2+s_y^3 s_x^4)
+c_x s_x^3 c_\d+c_\d c_y c_x s_x s_y-c_y s_x^2,\nnu\\
N_{\s1}&=&-s_\d (4 c_x^2 c_y c_\d^2 s_y s_x+2 c_x c_y^2 s_y c_\d-2 c_x^3 s_y c_y^2 c_\d
-2 c_\d c_x s_y
+4 c_\d c_x^3 s_y-c_x^2 s_x+s_y c_x^2 s_x c_y-c_y s_y s_x),\nnu\\
N_{\s2}&=& c_\d s_x+c_x c_y s_x^2+(1-2 s_\d^2) (s_y^3 c_x s_x^2
+ c_x s_y s_x^2-c_x s_y)
-c_\d s_x^3+(-4 s_\d^2 c_\d+3 c_\d) s_y s_x^3 c_y
+(4 s_\d^2 c_\d-2 c_\d) s_y s_x c_y,\nnu \\
\eea
while the corresponding expressions for $R_\n$, $M_{\b\b}$ and $M_\b$
are
\beq
R_\n\approx \left|{R_1\over R_2}\right|+O(s_z)
\eeq
where
\bea
R_1&=& -2 s_x c_x c_y c_\d s_y^2+2 s_y^2 s_x^2-s_y^2\nnu\\
R_2&=& 2 s_x^3 c_x c_y c_\d s_y^2-4 s_x^3 c_x c_\d s_y+2 s_x c_x c_\d s_y
+4 s_y c_y s_\d^2 s_x^4-2 s_y c_y s_x^4
-4 s_y c_y s_\d^2 s_x^2+2 s_y c_y s_x^2\nnu \\
&&-3 s_y^2 s_x^4+s_x^4+2 s_y^2 s_x^2-s_x^2,
\eea
\beq
M_{\b\b}\approx m_3\; {s_{2y} s_{2x}\over 4} \sqrt{{1\over
M_1}}+ O(s_z),
\eeq
where
\bea
M_1&=&s_y^4 s_x^4+3 s_x^4 s_y^2-s_x^4-3 s_x^2 s_y^2-s_y^4 s_x^2
+(4 c_\d^2-2) c_y s_y s_x^4+(2-4 c_\d ^2) c_y s_y
s_x^2+s_x^2+s_y^2\nnu \\
&&-4 c_\d s_x^3 c_x c_y s_y^2+4 c_\d s_y s_x^3 c_x-2 c_\d s_y s_x c_x
+2 s_y^2 c_y s_x c_x c_\d,
\eea
and
\beq
M_\b\approx m_3\; c_x \sqrt{{s_y^4 s_x^2-2 s_x^2 s_y^2\over M_2}} + O(s_z)
\eeq
where
\bea
M_2&=&2 c_\d s_y s_x c_x+s_x^4+(2-4 c_\d^2) c_y s_y s_x^4+3 s_x^2 s_y^2
-2 s_y^2 c_y s_x c_x c_\d+(4 c_\d^2-2) c_y s_y s_x^2\nnu \\
&&+4 c_\d s_x^3 c_x c_y s_y^2
-4 c_\d s_y s_x^3 c_x-s_x^2-s_y^2-s_y^4 s_x^4-3 s_x^4 s_y^2+s_y^4 s_x^2
\eea

\begin{figure}[hbtp]
\centering
\begin{minipage}[c]{0.5\textwidth}
\epsfxsize=6cm
\centerline{\epsfbox{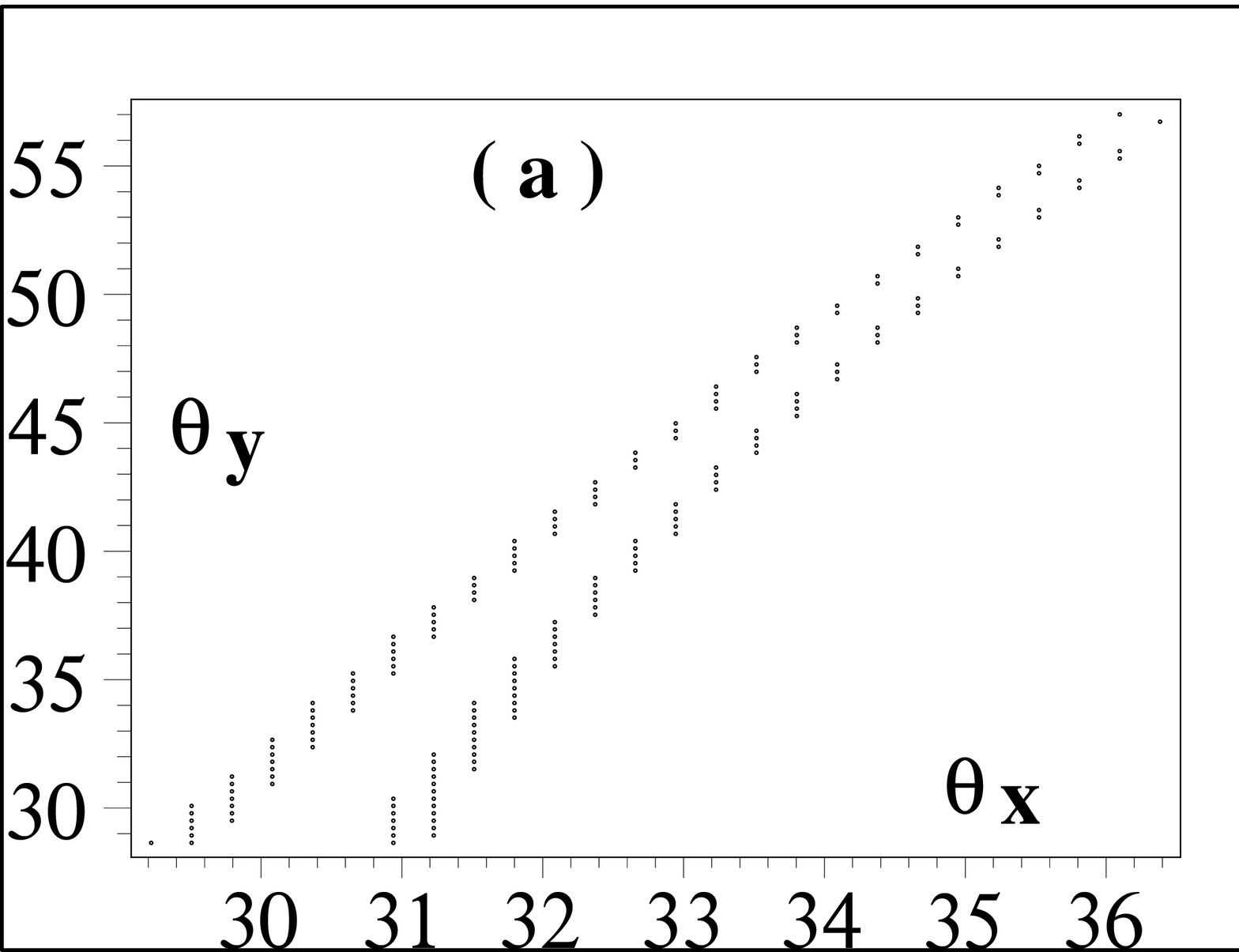}}
\end{minipage}%
\begin{minipage}[c]{0.5\textwidth}
\epsfxsize=6cm
\centerline{\epsfbox{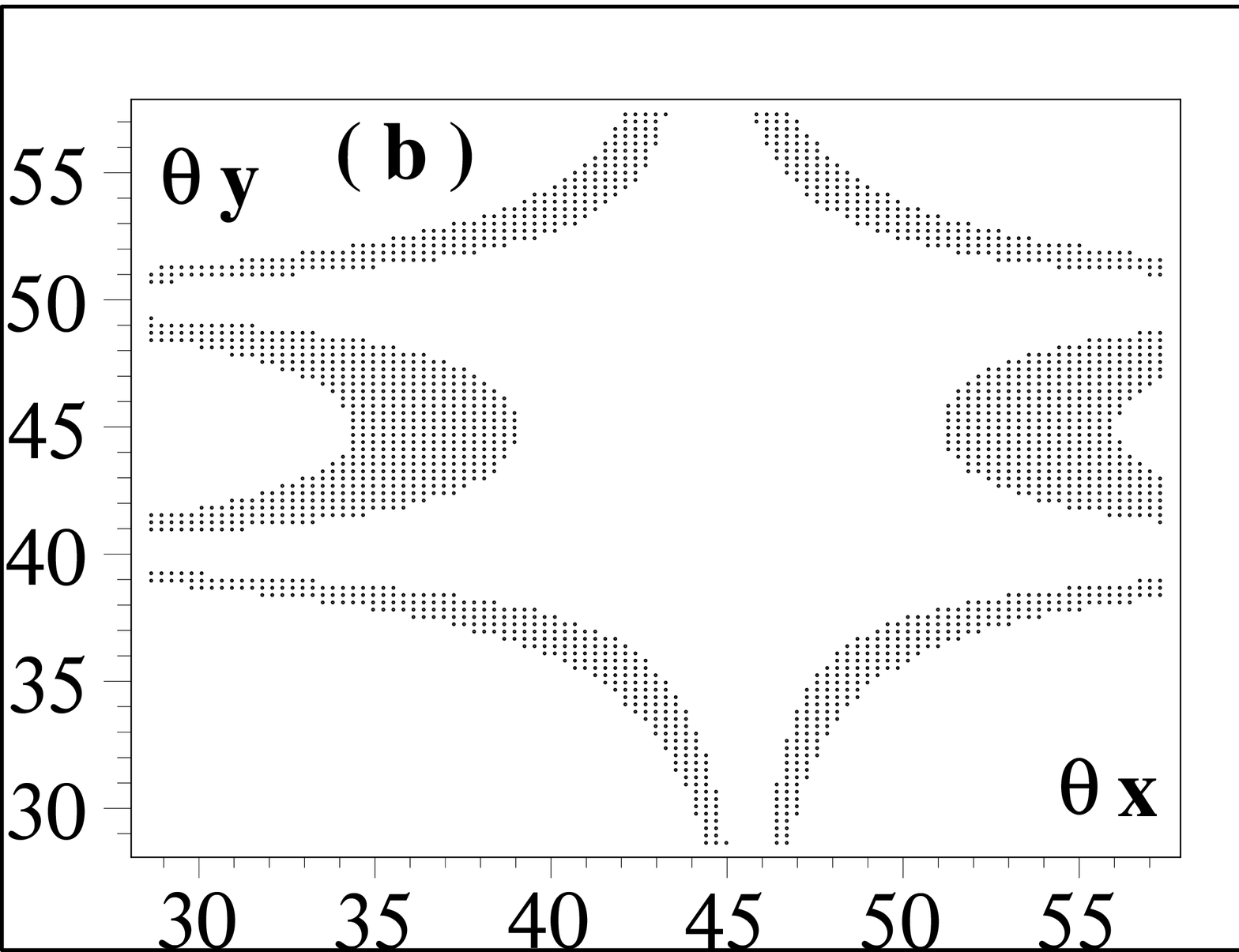}}
\end{minipage}
\caption{{\footnotesize The available parameter space for the cases
$\bf N_1$ (a) with $\d=138.1^0$and $\bf I_4$ (b) with $\d=31^0$ respectively.
The dotted region represents the set of points $(\t_x, \t_y)$ at
which $R_\n$ is constrained by $0.025 < R_\n <0.05$. The range from $29^0$ to
$57^0$ is spanned for both angles $\t_x$ and $\t_y$.  }}
\label{prfig}
\end{figure}

In this pattern, the required input values in order to match the experimental results
are $(\t_x = 33^0, \t_y = 42^0, \d = 138.1^0, \t_z = 5^0)$.
For these inputs we obtain $m_1/m_3 = 0.6563621852$, $m_2/m_3 =
0.6416046770$, $\r  = 79.56^0$, $\s = 148.87^0$ and $R_\nu = 0.033$.
The mass $m_3$ fitted from the observed $\D m^2_{\mbox{sol}}$ is $m_3 = 0.064\;
\mbox{eV}$. Then the derived values for the other remaining
parameters are $\D m^2_{\mbox{atm}}=2.4\times 10^{-3}\;\mbox{eV}^2$, $M_\b
=0.042\;\mbox{eV}$, $M_{\b\b} =0.021\;\mbox{eV}$ and $\Sigma =0.15 \;\mbox{eV}
$. As it is evident from Fig.~\ref{prfig}~(a), the available parameter space
permits the choice of $\t_x$ and $\t_y$ in the acceptable range without tuning.
In this pattern the numerically
estimated mass matrix $M$ is
\beq
M=
m_3\,\left(\begin {array}{ccc}
- 0.33257- 0.00386\,i&- 0.33256- 0.003873\,i& 0.45632+ 0.003944\,i\\
- 0.33256- 0.003873\,i&0.67411- 0.003791\,i& 0.33257+ 0.003870\,i\\
 0.45632+ 0.003944\,i& 0.33257+ 0.003870\,i& 0.64681- 0.003947\,i
\end {array}
\right)
\label{massa6}
\eeq

{\bf Pattern} $\bf N_2$: $M_{e\mu} + M_{\mu\tau} = 0, M_{\mu e} + M_{\tau\tau} =0.$
In this pattern the required quantities $a$'s and $b$'s as given
by \eq{abs} are
\bea
&& a_1=(-c_x s_y s_z-s_x c_y e^{-i\,\d}) c_x c_z+
(-c_x s_y s_z-s_x c_y e^{-i\,\d}) (-c_x c_y s_z+s_x s_y e^{-i\,\d}),\nnu\\
&& a_2=s_x c_z (-s_x s_y s_z+c_x c_y e^{-i\,\d})
+(-s_x s_y s_z+c_x c_y e^{-i\,\d}) (-s_x c_y s_z-c_x s_y e^{-i\,\d}), \nnu \\
&& a_3=s_y c_z (s_z+c_y c_z),\nnu \\
&& b_1=(-c_x s_y s_z-s_x c_y e^{-i\,\d}) c_x c_z+(-c_x c_y s_z
+s_x s_y e^{-i\,\d})^2,\nnu \\
&& b_2=(-s_x s_y s_z+c_x c_y e^{-i\,\d}) s_x c_z
+(-s_x c_y s_z-c_x s_y e^{-i\,\d})^2,\nnu \\
&& b_3=s_y s_z c_z+c_y^2 c_z^2 ,
\label{abn2}
\eea
which are sufficient to calculate all other quantities.
The corresponding expression for  $R_\n$, expanded at
the leading power of $s_z$, is
\beq
R_\n\approx \left|{4 c_x s_x^3 c_y s_y^6 c_\d-2 c_x s_x^3 c_y s_y^4 c_\d
-4 s_y^7 c_x s_x^3 c_\d+6 s_y^5 c_x s_x^3 c_\d
-2 s_y^3 c_x s_x^3 c_\d-4 s_y^5 c_y s_x^4+2 s_y^5 c_y s_x^2
-2 s_y^4 s_x^4+s_y^4 s_x^2\over
D_{N2}}\right|
+O(s_z).
\eeq
where
\bea
D_{N2} & =& -4 c_x s_x^5 c_y s_y^6 c_\d+2 c_x s_x^5 c_y s_y^4 c_\d
+4 s_y^7 c_x s_x^5 c_\d-6 s_y^5 c_x s_x^5 c_\d
+2 s_y^3 c_x s_x^5 c_\d\nnu \\
&&+6 s_y^5 c_y s_x^6-4 s_y^5 c_y s_x^4
-4 s_y^6 s_x^6+7 s_y^4 s_x^6-s_x^6 s_y^2+4 s_y^6 s_x^4-6 s_y^4 s_x^4+s_x^4 s_y^2.
\eea
In this pattern, matching the experimental results, we find that the required input
values are $(\t_x = 41.5^0, \t_y = 47^0, \d = 195^0, \t_z = 5^0)$.
For these inputs we obtain $m_1/m_3 = 0.7674926621$, $m_2/m_3 =
0.7584041167$, $\r  = 96.62^0$, $\s = 8.4^0$ and $R_\nu = 0.033$.
The mass $m_3$ fitted from the observed $\D m^2_{\mbox{sol}}$ is $m_3 = 0.076\;
\mbox{eV}$. Then the derived values for the other remaining
parameters are $\D m^2_{\mbox{atm}}=2.3\times 10^{-3}\;\mbox{eV}^2$, $M_\b
=0.058\;\mbox{eV}$, $M_{\b\b} =0.007\;\mbox{eV}$ and $\Sigma =0.19 \;\mbox{eV} $.
In this model $\t_x$ turns out to be a little bit out of the
allowed range.
In this pattern the numerically
estimated mass matrix $M$ is
\beq
M=
m_3\,\left(
\begin {array}{ccc}
- 0.092- 0.0024\,i&- 0.4444+ 0.000018\,i& 0.617+ 0.00029\,i\\
-0.444+ 0.000018\,i& 0.639+ 0.000014\,i& 0.444-0.000018\,i\\
 0.617+ 0.00029\,i&0.444- 0.000018\,i& 0.444- 0.000018\,i
 \end {array}
\right)
\label{massa7}
\eeq

\section{Inverted hierarchy models}
{\bf Pattern} $\bf I_1$: $M_{ee} + M_{\mu\mu} = 0, M_{ee} + M_{\mu\tau} =0.$
The $a$'s and $b$'s quantities can be inferred from the corresponding ones
of cases $\bf D_1$ and $\bf N_1$.
The corresponding expression for  $R_\n$, expanded at
the leading power of $s_z$, is
\beq
R_\n\approx \left|{2 c_y s_y^3-4 c_y s_y^3 s_x^2-s_y^4+2 s_y^4 s_x^2\over
D_{I1}}\right|
+O(s_z),
\eeq
where
\bea
D_{I1} &=&-1+(4+4 c_\d^2) c_y s_y^3 s_x^4-8 c_y s_y s_x^4+4 s_x^2+2 s_y^2
+(-3+4 c_\d^2) s_y^4 s_x^4+(-4 c_\d^2+8) s_x^4 s_y^2
-2 c_y s_y\nnu \\
&&-4 s_x^4+8 c_y s_y s_x^2+(-8+4 c_\d^2) s_x^2 s_y^2
+(-4 c_\d^2+2) s_y^4 s_x^2+(-2-4 c_\d^2) c_y s_y^3 s_x^2
\eea
In this pattern, to match the experimental results,
 the required input
values are $(\t_x = 43^0, \t_y = 44^0, \d = 26^0, \t_z = 5^0)$.
For these inputs we obtain $m_1/m_3 = 4.230294841$, $m_2/m_3 =
4.298049784$, $\r  = 103.57^0$, $\s = 15.37^0$ and $R_\nu = 0.033$.
The mass $m_3$ fitted from the observed $\D m^2_{\mbox{sol}}$ is $m_3 = 0.012\;
\mbox{eV}$. Then the derived values for the other remaining
parameters are $\D m^2_{\mbox{atm}}=2.3\times 10^{-3}\;\mbox{eV}^2$, $M_\b
=0.05\;\mbox{eV}$, $M_{\b\b} =0.003 \;\mbox{eV}$ and $\Sigma =0.11 \;\mbox{eV} $.
In this pattern the numerically
estimated mass matrix $M$ is
\beq
M=
m_3\,\left(
\begin {array}{ccc}
- 0.2982+ 0.0109\,i& 3.1198+ 0.17560\,i&- 2.8549- 0.17093\,i\\
 3.1198+ 0.17560\,i& 0.29818-0.01085\,i& 0.2982- 0.01090\,i\\
 - 2.8549- 0.17093\,i& 0.2982- 0.01090\,i& 1.0592+ 0.03131\,i
\end {array}
\right)
\label{massa2}
\eeq

{\bf Pattern} $\bf I_2$: $M_{ee} + M_{\mu\mu} = 0, M_{\mu e} + M_{\tau\tau} =0.$
The $a$'s and $b$'s quantities can be inferred from the corresponding ones
of cases $\bf D_1$ and $\bf N_2$.
The corresponding expression for  $R_\n$, expanded at
the leading power of $s_z$, is
\beq
R_\n\approx \left|{-6 s_x c_x c_y c_\d s_y^4+4 s_x c_x c_y c_\d s_y^2
+6 s_y^4 s_x^2-4 s_x^2 s_y^2-3 s_y^4+2 s_y^2\over
D_{I2}}\right|
+O(s_z),
\eeq
where
\bea
D_{I2} &=&-1+(4 c_\d^2-10) s_x^2 s_y^2+6 s_x^3 c_x c_y c_\d s_y^4
-8 s_x^3 c_x c_y c_\d s_y^2+(-4 c_\d^2+2) s_y^4 s_x^2\nnu\\
&&+(4 c_\d^2-5) s_y^4 s_x^4+(-4 c_\d^2+12) s_x^4 s_y^2
-4 s_x^4+2 s_y^2+4 s_x^2+2 s_x c_x c_y c_\d s_y^2
\eea
In this pattern, to match the experimental results, the required
values are $(\t_x = 34^0, \t_y = 43^0, \d = 118^0, \t_z = 5^0)$.
For these inputs we obtain $m_1/m_3 = 2.062421918$, $m_2/m_3 =
2.090403819$, $\r  = 77.05^0$, $\s = 132.41^0$ and $R_\nu = 0.034$.
The mass $m_3$ fitted from the observed $\D m^2_{\mbox{sol}}$ is $m_3 = 0.026\;
\mbox{eV}$. Then the derived values for the other remaining
parameters are $\D m^2_{\mbox{atm}}=2.2\times 10^{-3}\;\mbox{eV}^2$, $M_\b
=0.054\;\mbox{eV}$, $M_{\b\b} =0.034 \;\mbox{eV}$ and $\Sigma =0.135 \;\mbox{eV} $.
In this pattern the numerically
estimated mass matrix $M$ is
\beq
M=
m_3\,\left(
\begin {array}{ccc}
- 1.3165- 0.03164\,i&- 1.0155- 0.02210\,i& 1.2241+ 0.02439\,i\\
- 1.0155- 0.02210\,i& 1.3165+0.03163\,i&- 0.17371- 0.02687\,i\\
 1.2241+ 0.02439\,i&- 0.17371- 0.02687\,i& 1.0155+ 0.02213\,i
\end {array}
\right)
\label{massa3}
\eeq

{\bf Pattern} $\bf I_3$: $M_{e\mu} + M_{\mu\tau} = 0, M_{\mu\mu} + M_{\tau\tau} =0.$
The $a$'s and $b$'s quantities can be inferred from the corresponding ones
of cases $\bf D_3$ and $\bf N_1$.
The corresponding expression for  $R_\n$, expanded at
the leading power of $s_z$, is
\beq
R_\n\approx \left|{s_y c_\d s_x c_x+2 s_x^2 s_y^2-s_y^2\over
-s_y c_\d s_x^3 c_x-s_x^4 s_y^2}\right|
+O(s_z),
\eeq
In this pattern, to match the experimental results, the required
values are $(\t_x = 34^0, \t_y = 42^0, \d = 304^0, \t_z = 5^0)$.
For these inputs we obtain $m_1/m_3 = 1.676711363$, $m_2/m_3 =
1.694189580$, $\r  = 166.78^0$, $\s = 47.34^0$ and $R_\nu = 0.032$.
The mass $m_3$ fitted from the observed $\D m^2_{\mbox{sol}}$ is $m_3 = 0.037\;
\mbox{eV}$. Then the derived values for the other remaining
parameters are $\D m^2_{\mbox{atm}}=2.4\times 10^{-3}\;\mbox{eV}^2$, $M_\b
=0.062\;\mbox{eV}$, $M_{\b\b} =0.036\;\mbox{eV}$ and $\Sigma =0.160 \;\mbox{eV} $.
In this pattern the numerically
estimated mass matrix $M$ is
\beq
M=
m_3\,\left(
\begin {array}{ccc}
0.98880+ 0.01487\,i&- 1.0068- 0.00011\,i& 0.90785- 0.00166\,i\\
- 1.0068- 0.00011\,i& 0.01346-0.00008\,i& 1.0068+ 0.00010\,i\\
0.90785- 0.00166\,i& 1.0068+ 0.00010\,i&- 0.01346+ 0.00010\,i
\end {array}
\right)
\label{massa5}
\eeq

{\bf Pattern} $\bf I_4 $: $M_{\mu\mu} + M_{\tau\tau} = 0, M_{e e} + M_{\mu\tau} =0.$
In this pattern the required quantities $a$'s and $b$'s as given
by \eq{abs} are
\bea
a_1=c_x^2 s_z^2 + s_x^2 e^{-i\,2\,\d}, & a_2 =s_x^2 s_z^2 + c_x^2
e^{-i\,2\,\d}, & a_3=c_z^2, \nnu\\
\eea
where the coefficients $b$'s have the values as given in
\eq{abn1}.

Using $s_z$ as a small parameter, expanding in terms of its power and keeping
only leading terms,
we have the analytical approximate formulae for the mass ratios:
\bea
{m_1\over m_3} &\approx&c_{2x}^{-1} \sqrt{1-2 c_x^2-4 c_x^4 s_y c_y+4 c_x^2 s_y c_y
+4 c_x^4 c_y^2-4 c_x^4 c_y^4-8 c_x^2 s_y c_y c_\d^2
+c_x^4+8 c_\d^2 c_x^4 s_y c_y} + O(s_z)\nnu \\
{m_2\over m_3} &\approx& c_{2x}^{-1} \sqrt{4 c_y^2-4 c_x^4 s_y c_y-4 c_y^4
+4 c_x^2 s_y c_y-8 c_x^2 c_y^2+8 c_x^2 c_y^4+4 c_x^4 c_y^2-4 c_x^4 c_y^4
-8 c_x^2 s_y c_y c_\d^2+c_x^4+8 c_\d^2 c_x^4 s_y c_y} +
O(s_z)\nnu\\
\eea

The corresponding expression for $\r$ and $\s$ are
\bea
\r \approx {1\over 2} \tan^{-1}\left({s_x^2 s_{2\d} \over s_x^2 c_{2\d} - s_{2y} c_x^2}\right) +
O(s_z),
&&
\s \approx {1\over 2}  \tan^{-1}\left({ -c_x^2 s_{2\d}
\over s_{2y} s_x^2-c_x^2 c_{2\d}}\right) +
O(s_z),
\eea

While the corresponding expression for $R_\n$, $M_{\b\b}$ and $M_\b$
are
\beq
R_\n\approx \left|{1+4 s_y^4-8 s_y^4 s_x^2-4 s_y^2+8 s_x^2 s_y^2-2 s_x^2\over
(-8 c_\d^2+4) s_y c_y s_x^4-2 s_x^2+3 s_x^4
+(-4+8 c_\d^2) s_y c_y s_x^2+4 s_y^4 s_x^4-4 s_x^4 s_y^2}\right|+O(s_z),
\eeq
\beq
M_{\b\b}\approx m_3\;s_{2y} + O(z),\;\;\; M_\b \approx m_3\;c_{2x}^{-1}
\sqrt{M3},
\eeq
where
\beq
M_3 = 8 c_\d^2 c_x^4 s_y c_y-8 c_x^2 s_y c_y c_\d^2
-12 c_x^2 c_y^2+4 c_x^2 s_y c_y-4 c_x^4 s_y c_y
+12 c_x^2 c_y^4+12 c_x^4 c_y^2
-12 c_x^4 c_y^4+c_x^2-c_x^4+4 c_y^2-4 c_y^4
\eeq

In this pattern, no need for  tuning   to match the
experimental results as it is clear from the parameter space in Fig.~\ref{prfig}(b).
The required
values are $(\t_x = 34^0, \t_y = 42^0, \d = 31^0, \t_z = 5^0)$.
For these inputs we obtain $m_1/m_3 = 1.601974769$, $m_2/m_3 =
1.585960893$, $\r  = 76.95^0$, $\s = 134.67^0$ and $R_\nu = 0.034$.
The mass $m_3$ fitted from the observed $\D m^2_{\mbox{sol}}$ is $m_3 = 0.039\;
\mbox{eV}$. Then the derived values for the other remaining
parameters are $\D m^2_{\mbox{atm}}=2.4\times 10^{-3}\;\mbox{eV}^2$, $M_\b
=0.063\;\mbox{eV}$, $M_{\b\b} =0.039\;\mbox{eV}$ and $\Sigma =0.165 \;\mbox{eV} $.
In this pattern the numerically
estimated mass matrix $M$ is
\beq
M=
m_3\,\left(
\begin {array}{ccc} - 0.97926- 0.01129\,i& 0.12874- 0.92428\,i& 0.11713+ 0.83357\,i\\
 0.12874- 0.92428\,i&-0.10446+ 0.10832\,i& 0.97931+ 0.01130\,i\\
 0.11713+0.83357\,i& 0.97931+ 0.01130\,i& 0.10442- 0.10834\,i
 \end {array}
\right)
\label{massa8}
\eeq

{\bf Pattern} $\bf I_5$: $M_{ee} + M_{\tau\tau} = 0, M_{e e} + M_{\mu\tau} =0.$
The $a$'s quantities can be inferred from the corresponding ones
of cases $\bf D_2$, while $b$'s are given as
\bea
b_1=c_x c_z (c_x c_z-c_x c_y s_z+s_x s_y e^{-i\,\d}),&
b_2=-s_x c_z (-s_x c_z+s_x c_y s_z+c_x s_y e^{-i\,\d}), &
b_3 = s_z (s_z+c_y c_z).
\label{abi5}
\eea
The corresponding expression for  $R_\n$, expanded at
the leading power of $s_z$, is
\beq
R_\n\approx \left|{-c_y^3(-c_y+2 c_y c_x^2+2 s_y-4 s_y c_x^2)\over
D_{I5}}\right|
+O(s_z),
\eeq
where
\bea
D_{I5}&=&-1+4 c_x^2+2 c_y^2+4 c_y^2 c_x^2 c_\d^2-4 c_\d^2 c_y^3 c_x^2 s_y
+4 c_y^3 c_x^4 c_\d^2 s_y+2 c_y^3 s_y-6 c_y^3 c_x^2 s_y-8 c_y s_y c_x^4
-8 c_y^2 c_x^2\nnu \\
&&-c_y^4-2 c_y s_y-3 c_y^4 c_x^4-4 c_x^4-4 c_y^4 c_x^2 c_\d^2
+4 c_y^4 c_x^4 c_\d^2-4 c_y^2 c_x^4 c_\d^2+8 c_x^4 c_y^2
+8 c_y s_y c_x^2+4 c_y^4 c_x^2+4 c_y^3 c_x^4 s_y\nnu\\
\eea
In this pattern, there is no  tuning for the angles $\t_x$  to match the
experimental results. The required
values are $(\t_x = 44.73^0, \t_y = 42^0, \d = 90^0, \t_z = 5^0)$.
For these inputs we obtain $m_1/m_3 = 4.137879853$, $m_2/m_3 =
4.076054427$, $\r  = 51.24^0$, $\s = 136.24^0$ and $R_\nu = 0.033$.
The mass $m_3$ fitted from the observed $\D m^2_{\mbox{sol}}$ is $m_3 = 0.012\;
\mbox{eV}$. Then the derived values for the other remaining
parameters are $\D m^2_{\mbox{atm}}=2.5\times 10^{-3}\;\mbox{eV}^2$, $M_\b
=0.051\;\mbox{eV}$, $M_{\b\b} =0.004\;\mbox{eV}$ and $\Sigma =0.115 \;\mbox{eV} $.
In this pattern the numerically
estimated mass matrix $M$ is
\beq
M=
m_3\,\left(
\begin {array}{ccc}
- 0.35325+ 0.0220\,i&- 2.9235- 0.39769\,i& 2.7916+ 0.35549\,i\\
- 2.9235- 0.39769\,i& 0.98990+0.0765\,i& 0.35326- 0.0220\,i\\
 2.7916+ 0.35549\,i& 0.35326- 0.0220\,i& 0.35325- 0.02201\,i
 \end {array}
\right)
\label{massa10}
\eeq

{\bf Pattern} $\bf I_6$: $M_{ee} + M_{\mu\mu} = 0, M_{\mu\mu} + M_{\tau\tau}
=0$.
The $a$'s and $b$'s quantities can be inferred from the corresponding ones
of cases $\bf D_2$ and $\bf D_3$.
The corresponding expression for  $R_\n$, expanded at
the leading power of $s_z$, is
\beq
R_\n\approx \left|{s_y^4-2 s_y^4 s_x^2-s_y^2+2 s_x^2 s_y^2\over
s_y^4 s_x^4+2 s_\d^2 s_y^2 s_x^2-2 s_\d^2 s_y^2 s_x^4-s_x^2 s_\d^2
+s_x^4 s_\d^2+s_x^2-s_x^4-s_x^2 s_y^2}\right|
+O(s_z).
\eeq
In this pattern, no  tuning is needed  to match the
experimental results. The required
values are $(\t_x = 42^0, \t_y = 42^0, \d = 170^0, \t_z = 5^0)$.
For these inputs we obtain $m_1/m_3 = 8.699398197$, $m_2/m_3 =
8.840977186$, $\r  = 174.85^0$, $\s = 83.86^0$ and $R_\nu = 0.032$.
The mass $m_3$ fitted from the observed $\D m^2_{\mbox{sol}}$ is $m_3 = 0.006\;
\mbox{eV}$. Then the derived values for the other remaining
parameters are $\D m^2_{\mbox{atm}}=2.4\times 10^{-3}\;\mbox{eV}^2$, $M_\b
=0.049\;\mbox{eV}$, $M_{\b\b} =0.005\;\mbox{eV}$ and $\Sigma =0.105 \;\mbox{eV} $.
In this pattern the numerically
estimated mass matrix $M$ is
\beq
M=
m_3\,\left(
\begin {array}{ccc}
0.8601- 0.01705\,i& 6.4626- 0.14461\,i&- 5.8026+ 0.13222\,i\\
 6.4626- 0.14461\,i&- 0.8601+0.01705\,i& 0.9140+ 0.00166\,i\\
 - 5.8026+ 0.13222\,i& 0.9140+ 0.00166\,i& 0.8602- 0.01705\,i
\end {array}
\right)
\label{massb1}
\eeq

{\bf Pattern} $\bf I_7$: $M_{ee} + M_{\tau\tau} = 0,\; M_{e \mu} + M_{\tau\tau} =0.$
The $a$'s and $b$'s quantities can be inferred from the corresponding ones
of cases $\bf D_2$ and $\bf D_3$.
The corresponding expression for  $R_\n$, expanded at
the leading power of $s_z$, is
\beq
R_\n\approx \left|{c_y^4 (2 c_\d s_x c_x c_y-1+2 c_x^2)\over
D_{I7}}\right|
+O(s_z),
\eeq
where
\bea
D_{I7}&=&1+4 c_x^2 c_y^2 c_\d^2-4 c_\d^2 c_x^4 c_y^2-4 c_x^2+4 c_y^4 c_x^4 c_\d^2
+8 c_x^2 c_y^2+6 c_\d s_x c_x c_y^3-4 c_\d s_x c_x c_y
-4 c_y^4 c_x^2 c_\d^2+4 c_x^4+c_y^4\nnu \\
&&-12 c_\d c_x^3 c_y^3 s_x
-2 c_y^2+8 c_\d c_x^3 c_y s_x+3 c_y^4 c_x^4-4 c_y^4 c_x^2-8 c_x^4 c_y^2
-2 c_y^5 c_\d s_x c_x+2 c_x^3 c_y^5 c_\d s_x
\eea
In this pattern, no tuning for the angles $\t_x$  to match the
experimental results. The required
values are $(\t_x = 31^0, \t_y = 44^0, \d = 147^0, \t_z = 5^0)$.
For these inputs we obtain $m_1/m_3 = 1.475002097$, $m_2/m_3 =
1.462135796$, $\r  = 81.84^0$, $\s = 154.74^0$ and $R_\nu = 0.033$.
The mass $m_3$ fitted from the observed $\D m^2_{\mbox{sol}}$ is $m_3 = 0.046\;
\mbox{eV}$. Then the derived values for the other remaining
parameters are $\D m^2_{\mbox{atm}}=2.5\times 10^{-3}\;\mbox{eV}^2$, $M_\b
=0.067\;\mbox{eV}$, $M_{\b\b} =0.036\;\mbox{eV}$ and $\Sigma =0.180 \;\mbox{eV} $.
In this pattern the numerically
estimated mass matrix $M$ is
\beq
M=
m_3\,\left(
\begin {array}{ccc}
- 0.77989+ 0.00498\,i&- 0.77991+ 0.00497\,i& 0.96962- 0.00542\,i\\
- 0.77991+ 0.00497\,i&0.98867- 0.00664\,i& 0.10578+ 0.00582\,i\\
0.96962-0.00542\,i& 0.10578+ 0.00582\,i& 0.77989- 0.00498\,i
\end {array}
\right)
\label{massb2}
\eeq

{\bf Pattern} $\bf I_8$: $M_{\mu\mu} + M_{\tau\tau} = 0, M_{ee} + M_{\tau\tau} =0.$
The $a$'s and $b$'s parameters can be inferred from the corresponding ones
of cases $\bf D_2$ and $\bf D_3$.
The corresponding expression for  $R_\n$, expanded at
the leading power of $s_z$, is
\beq
R_\n\approx \left|{4 s_y^4-8 s_y^4 s_x^2-4 s_y^2+8 s_x^2 s_y^2\over
4 s_y^4 s_x^4-8 s_\d^2 s_y^2 s_x^2+8 s_\d^2 s_y^2 s_x^4+4 s_x^2 s_y^2
-8 s_x^4 s_y^2+4 s_x^2 s_\d^2-4 s_x^4 s_\d^2}\right|
+O(s_z).
\eeq
In this pattern, no need for  tuning for the angles $\t_x$  to match the
experimental results. The acceptable
values are $(\t_x = 42.9^0, \t_y = 42^0, \d = 33^0, \t_z = 5^0)$.
For these inputs we obtain $m_1/m_3 = 10.88384938$, $m_2/m_3 =
11.06236809$, $\r  = 18.5^0$, $\s = 110.84^0$ and $R_\nu = 0.032$.
The mass $m_3$ fitted from the observed $\D m^2_{\mbox{sol}}$ is $m_3 = 0.004\;
\mbox{eV}$. Then the derived values for the other remaining
parameters are $\D m^2_{\mbox{atm}}=2.4\times 10^{-3}\;\mbox{eV}^2$, $M_\b
=0.049\;\mbox{eV}$, $M_{\b\b} =0.004\;\mbox{eV}$ and $\Sigma =0.103 \;\mbox{eV} $.
In this pattern the numerically
estimated mass matrix $M$ is
\beq
M=
m_3\,\left(
\begin {array}{ccc}
0.8378+ 0.1050\,i&- 8.0358- 0.90178\,i& 7.2547+ 0.79963\,i\\
- 8.0358- 0.90178\,i& 0.8375+0.1049\,i& 1.0923+ 0.0118\,i\\
 7.2547+ 0.79963\,i&1.0923+ 0.0118\,i&- 0.8376- 0.1048\,i
\end {array}
\right)
\label{massb3}
\eeq

The remaining two models of  $M_{11}+M_{33}=0,\; M_{12}+M_{23}=0$
and $M_{11}+M_{23}=0,\; M_{21}+M_{33}=0$ fail to be consistent
with the experimental data for any reasonable choice of the two
mixing angle $\t_x$ and $\t_y$.
\textheight      240mm  
{\footnotesize
\begin{landscape}
\begin{table}[h]
\begin{center}
\begin{tabular}{ |c||c||c|c|c|c|c|c|c|c|c|c|c|c|c|c|c|}
\hline
   \mbox{Model} & \mbox{Trace conditions}&\mbox{Status}   & $\t_x$
   & $\t_y$ & $\t_z$  & $\d$ & $R_\n$ & $\frac{m_1}{m_3}$ & $\frac{m_2}{m_3}$ &
   $\rho$& $\sigma$ &$m_3$&$M_{\b}$ & $M_{\b\b}$ & $\Sigma$ & $\D m_{\mbox{atm}}^2$\\
\hline \hline
${\bf D_1}$ & $(11,22),(12,23)$ & \mbox{allowed} & $34$ &  $42$ & $5$   & $92.755$ & $0.033$
& $1.050$ & $1.049$ & $87.72$ & $95$ &$0.156$& $0.164$& $0.160$& $0.484$
& $0.0025$ \\
\hline
${\bf D_2}$ & $(11,33),(11,22)$&\mbox{allowed} & $34$ &  $44.65$ & $5$   & $90$ & $0.035$
& $1.0152$ & $1.0147$ & $89.99$ & $89.86$ &$0.276$& $0.281$& $0.277$& $0.84$
& $0.0024$\\
\hline
${\bf D_3}$ & $(22,33),(21,33)$&\mbox{allowed} & $34$ &  $42$ & $5$   & $272.9$ & $0.034$
& $1.0081$ & $1.0079$ & $177.55$ & $5.32$ &$0.38$& $0.38$& $0.38$& $1.15$
& $0.0024$ \\
\hline
${\bf N_1}$ & $(12,23),(11,23)$&\mbox{allowed} & $33$ &  $42$ & $5$   & $138.1$ & $0.033$
& $0.656$ & $0.642$ & $79.56$ & $148.87$ &$0.064$& $0.042$& $0.021$& $0.148$
& $0.0024$ \\
\hline
${\bf N_2}$ & $(12,23),(12,33)$ &\mbox{disallowed} & $41.5$ &  $47$ & $5$   & $195$ & $0.033$
& $0.767$ & $0.758$ & $96.62$ & $8.4$ &$0.076$& $0.058$& $0.007$& $0.191$
& $0.0023$ \\
\hline
${\bf I_1}$ & $(11,22),(11,23)$ &\mbox{disallowed} &$43$ &  $44$ & $5$   & $26$ & $0.033$
& $4.230$ & $4.298$ & $103.57$ & $15.37$ &$0.012$& $0.05$& $0.003$& $0.111$
& $0.0023$ \\
\hline
${\bf I_2}$ & $(11,22),(21,33)$ &\mbox{allowed} & $34$ &  $43$ & $5$   & $118$ & $0.034$
& $2.0602$ & $2.0904$ & $77.05$ & $132.41$ &$0.026$& $0.054$& $0.034$& $0.135$
& $0.0022$ \\
\hline
${\bf I_3}$ & $(12,23),(22,33)$ &\mbox{allowed} & $34$ &  $42$ & $5$   & $304$ & $0.032$
& $1.677$ & $1.694$ & $166.78$ & $47.34$ &$0.037$& $0.062$& $0.036$& $0.160$
& $0.0024$ \\
\hline
${\bf I_4}$ & $(22,33),(11,23)$ & \mbox{allowed} & $34$ &  $42$ & $5$   & $31$ & $0.034$
& $1.6019$ & $1.5859$ & $76.95$ & $134.67$ &$0.039$& $0.063$& $0.039$& $0.165$
& $0.0024$ \\
\hline
${\bf I_5}$ & $(11,33),(23,11)$ &\mbox{disallowed} & $44.73$ &  $42$ & $5$   & $90$ & $0.033$
& $4.137$ & $4.076$ & $51.24$ & $136.24$ &$0.012$& $0.051$& $0.004$& $0.115$
& $0.0025$\\
\hline
${\bf I_6}$ & $(11,22),(22,33)$ &\mbox{disallowed} & $42$ &  $42$ & $5$   & $170$ & $0.032$
& $8.699$ & $8.841$ & $174.85$ & $83.86$ &$0.006$& $0.050$& $0.005$& $0.105$
& $0.0024$ \\
\hline
${\bf I_7}$ & $(11,33),(21,33)$ &\mbox{allowed} & $31$ &  $44$ & $5$   & $147$ & $0.033$
& $1.475$ & $1.462$ & $81.84$ & $154.74$ &$0.046$& $0.067$& $0.036$& $0.180$
& $0.0025$\\
\hline
${\bf I_8}$ & $(11,33),(22,33)$ &\mbox{disallowed} & $42.9$ &  $42$ & $5$   & $33$ & $0.032$
& $10.88$ & $11.062$ & $18.5$ & $110.84$ &$0.004$& $0.049$& $0.004$& $0.103$
& $0.0024$\\
\hline
\end{tabular}
\end{center}
 \caption{\small  The $13$ patterns for the
 two-vanishing traces. The trace corresponding to the index $(ab,ij)$
 is $M_{ab}+M_{ij}=0$. All the angles are
 measured in degrees, masses in eV and $\D m_{\mbox{atm}}^2$ in $\mbox{eV}^2$}
\label{tab1}
 \end{table}
\end{landscape}
}
\textheight      240mm  
{\footnotesize
\begin{table}[h]
\begin{center}
\begin{tabular}{|c||c|c|}
\hline
   $\mbox{Model}$ & $\mbox{Trace conditions}$& $M$\\
\hline \hline
${\bf D_1}$ &$(11,22), (12,23)$&
$m_3\left[
\begin {array}{ccc} - 1.0268+ 0.000489\,i& 0.027333- 0.00049\,i& 0.21400+ 0.00037\,i\\\noalign{\medskip} 0.027333- 0.00049\,i&
 1.0268- 0.000513\,i&- 0.02737+ 0.000517\,i\\\noalign{\medskip}
 0.21400+ 0.00037\,i&- 0.02737+ 0.000517\,i& 0.99942- 0.000507\,i
\end {array} \right]
$ \\
\hline ${\bf D_2}$   &$(11,33), (11,22)$&
$m_3\,\left[
\begin {array}{ccc}
- 0.99982+ 0.0017117\,i& 0.12446- 0.00027\,i& 0.12299+ 0.00004\,i\\\noalign{\medskip}
 0.12446- 0.00027\,i& 0.99981- 0.001714\,i&- 0.01515+ 0.0017276\,i\\\noalign{\medskip}
 0.12299+ 0.00004\,i&- 0.01515+ 0.0017276\,i& 0.99982- 0.001714\,i
\end {array}
\right]
$ \\
\hline ${\bf D_3}$   &$(22,33), (21,33)$&
$ m_3\,\left[
\begin {array}{ccc}  1.0001- 0.000939\,i&- 0.093575- 0.00001\,i& 0.084241+ 0.00011\,i\\\noalign{\medskip}- 0.093575- 0.00001\,i&-
 0.09358+ 0.000006\,i& 0.99567- 0.000001\,i\\\noalign{\medskip}
 0.084241+ 0.00011\,i& 0.99567- 0.000001\,i& 0.09355- 0.000011\,i
\end {array}
\right]$
\\
\hline ${\bf N_1}$ & $(12,23), (11,23)$& $m_3\,\left[
\begin{array}{ccc}
- 0.33257- 0.00386\,i&- 0.33256- 0.003873\,i& 0.45632+ 0.003944\,i\\\noalign{\medskip}
- 0.33256- 0.003873\,i&0.67411- 0.003791\,i& 0.33257+ 0.003870\,i\\\noalign{\medskip}
 0.45632+ 0.003944\,i& 0.33257+ 0.003870\,i&0.64681-0.003947\,i
\end {array}
\right]$\\
\hline
 ${\bf N_2}$ &$(12,23), (12,33)$& $m_3\,\left[
\begin {array}{ccc}
- 0.092- 0.0024\,i&- 0.4444+ 0.000018\,i& 0.617+ 0.00029\,i\\\noalign{\medskip}
-0.444+ 0.000018\,i& 0.639+ 0.000014\,i& 0.444-0.000018\,i\\\noalign{\medskip}
 0.617+ 0.00029\,i&0.444- 0.000018\,i& 0.444- 0.000018\,i
 \end {array}
\right]
$ \\
\hline ${\bf I_1}$    &$(11,22), (11,23)$&
$m_3\,\left[
\begin {array}{ccc}
- 0.2982+ 0.0109\,i& 3.1198+ 0.17560\,i&- 2.8549- 0.17093\,i\\\noalign{\medskip}
 3.1198+ 0.17560\,i& 0.29818-0.01085\,i& 0.2982- 0.01090\,i\\\noalign{\medskip}
 - 2.8549- 0.17093\,i& 0.2982- 0.01090\,i& 1.0592+ 0.03131\,i
\end {array}
\right]
$ \\
\hline ${\bf I_2}$    &$(11,22), (21,33)$&
$m_3\,\left[
\begin {array}{ccc}
- 1.3165- 0.03164\,i&- 1.0155- 0.02210\,i& 1.2241+ 0.02439\,i\\\noalign{\medskip}
- 1.0155- 0.02210\,i& 1.3165+0.03163\,i&- 0.17371- 0.02687\,i\\\noalign{\medskip}
 1.2241+ 0.02439\,i&- 0.17371- 0.02687\,i& 1.0155+ 0.02213\,i
\end {array}
\right]$ \\
\hline ${\bf I_3}$  &$(12,23), (22,33)$&
$m_3\,\left[
\begin {array}{ccc}
0.98880+ 0.01487\,i&- 1.0068- 0.00011\,i& 0.90785- 0.00166\,i\\\noalign{\medskip}
- 1.0068- 0.00011\,i& 0.01346-0.00008\,i& 1.0068+ 0.00010\,i\\\noalign{\medskip}
0.90785- 0.00166\,i& 1.0068+ 0.00010\,i&- 0.01346+ 0.00010\,i
\end {array}
\right]$ \\
\hline ${\bf I_4}$  &$(22,33), (11,23)$&
$m_3\,\left[
\begin {array}{ccc} - 0.97926- 0.01129\,i& 0.12874- 0.92428\,i& 0.11713+ 0.83357\,i\\\noalign{\medskip}
 0.12874- 0.92428\,i&-0.10446+ 0.10832\,i& 0.97931+ 0.01130\,i\\\noalign{\medskip}
 0.11713+0.83357\,i& 0.97931+ 0.01130\,i& 0.10442- 0.10834\,i
 \end {array}
\right]
$ \\
\hline ${\bf I_5}$  &$(11,33), (23,11)$&
$m_3\,\left[
\begin {array}{ccc}
- 0.35325+ 0.0220\,i&- 2.9235- 0.39769\,i& 2.7916+ 0.35549\,i\\\noalign{\medskip}
- 2.9235- 0.39769\,i& 0.98990+0.0765\,i& 0.35326- 0.0220\,i\\\noalign{\medskip}
 2.7916+ 0.35549\,i& 0.35326- 0.0220\,i& 0.35325- 0.02201\,i
 \end {array}
\right]
$\\
\hline ${\bf I_6}$  &$(11,22), (22,33)$&
$m_3\,\left[
\begin {array}{ccc}
0.8601- 0.01705\,i& 6.4626- 0.14461\,i&- 5.8026+ 0.13222\,i\\\noalign{\medskip}
 6.4626- 0.14461\,i&- 0.8601+0.01705\,i& 0.9140+ 0.00166\,i\\\noalign{\medskip}
 - 5.8026+ 0.13222\,i& 0.9140+ 0.00166\,i& 0.8602- 0.01705\,i
\end {array}
\right]
$\\
\hline ${\bf I_7}$  &$(11,33), (21,33)$&
$m_3\,\left[
\begin {array}{ccc}
- 0.77989+ 0.00498\,i&- 0.77991+ 0.00497\,i& 0.96962- 0.00542\,i\\\noalign{\medskip}
- 0.77991+ 0.00497\,i&0.98867- 0.00664\,i& 0.10578+ 0.00582\,i\\\noalign{\medskip}
0.96962-0.00542\,i& 0.10578+ 0.00582\,i& 0.77989- 0.00498\,i
\end {array}
\right]
$\\
\hline ${\bf I_8}$  &$(11,33), (22,33)$&
$m_3\,\left[
\begin {array}{ccc}
0.8378+ 0.1050\,i&- 8.0358- 0.90178\,i& 7.2547+ 0.79963\,i\\\noalign{\medskip}
- 8.0358- 0.90178\,i& 0.8375+0.1049\,i& 1.0923+ 0.0118\,i\\\noalign{\medskip}
7.2547+ 0.79963\,i&1.0923+ 0.0118\,i&- 0.8376- 0.1048\,i
\end {array}
\right]
$\\
\hline
\end{tabular}
\end{center}
 \caption{\small  The mass estimates for the $13$ acceptable patterns.
 The trace corresponding to the index $(ab,ij)$ is~$M_{ab}~+~M_{ij}~=~0$.}
\label{tab2}
 \end{table}
}
\section{Conclusion and discussion}
In this work several patterns of Majorana neutrino mass matrix,
consistent with the present available observed data, are derived.
The models are based on textures possessing two $2\times 2$
sub-matrices with vanishing trace. The new proposed texture can
be considered as a non trivial generalization of the zero-texture
as explained in the introduction.

In our work we have thirteen possible acceptable patterns for Majorona mass
matrix  out of fifteen ones. The resulting models fall into three
distinct classes namely, degenerate case
($\bf D_1\cdots D_3$), normal hierarchy case ($\bf N_1\, \mbox{and}\, N_2$),
and inverted  hierarchy case ($\bf I_1\cdots I_8$). The  numerical
results of our study is summarized in Table~(\ref{tab1})and
Table~(\ref{tab2}) for a quick reference.

Our numerical study reveals that there are eight models ($\bf D_1, D_2, D_3, N_1, I_2,
I_3, I_4\, \mbox{and}\, I_7$), for which the mixing angles ($\t_x, \t_y\, \mbox{and}
\,\t_z$) can be adjusted to fall into the acceptable range given
by~\eq{osp}. In the remaining five models,($\bf N_2,  I_1,
I_5, I_6\, \mbox{and}\, I_8$),  $\t_x$ falls in the range $41.5^0 \le \t_x\le 45^0$,
which is out of the acceptable range.  These kinds of models can be considered as
empirically ruled out. An avenue for curing these models could be provided by a
small perturbation over the adopted textures. In our subsequent discussion we
only focus on the successful models namely ($\bf D_1, D_2, D_3, N_1, I_2,
I_3, I_4\, \mbox{and}\, I_7$).

The numerical study points out that Dirac-type phase $\d$\footnote{It is numerically
observed  that if there is an acceptable value for $\d$  say $\d_1$, then there is
another one $\d_2$ such that $\d_1 + \,\d_2 = 2\,\pi$} tends to be around ${\pi\over 2}
\;\mbox{or}\; {3\,\pi\over 2}$ in the degenerate case, while in other cases no general trend
could be observed. Concerning a possible relation between Dirac
and Majorana phases that could be revealed by numerical study, we
find the two phases $\r$ and $\d$  almost satisfying the
relation $\r\approx {\d\over 2}$ in the normal hierarchy case,
while for the degenerate case there is the relation
$\r\approx\d\; \mbox{or} \;\d-{\pi\over 2}$,
and for the inverted hierarchy case $\r \approx {\d\over 2}$ is
obeyed except for $\bf I_4$ where $\r\approx 2\,\d$ is
satisfied.

Another possible relation between $\s$ and $\d$ could be easily
recognized. In the normal hierarchy and degenerate case, the
relation $\s\approx \d\; \mbox{or}\; \d-{3\,\pi\over 2} $ is satisfied.
The inverted hierarchy cases have no specific general relation which is obeyed.

The non oscillation parameters $M_\b , M_{\b\b}\, \mbox{and}\, \Sigma $ are
consistent with the bounds given in \eq{neq}. In all the successful
models, $M_{\b}$, $M_{\b\b}$ and $m_3$ have the same order of magnitude.
The mass sum parameter is always constrained to be  $\Sigma \le
1.15\;\mbox{eV}$ which is safe with the cosmological bound in \eq{neq}.

All successful models are found to be still consistent with experimental data
in the limit of vanishing $\t_z$ while keeping $\t_x\approx 34^0$ and
$\t_y\approx 42^0$ constants. The same
thing still holds, when $\t_z$ is stretched  to its upper bound $(10^0)$.
In these limits, little changes take place for the other parameters.

Regarding the hierarchical structure of the mass matrices, as it
is evident from table.~\ref{tab2}, all successful models have clear hierarchical
structure except the mass matrix of model $\bf N_1$  whose
elements have all the same order of magnitude. These hierarchical
properties are restricted to the real parts, but for the imaginary
part they are always very small in comparison with the real ones with the exception
of model $\bf I_4$.

Final remark, related to when we restrict the study to the parameter space
$(\t_x\approx 34^0, \t_y\approx 42^0\;\mbox{and}\; \t_z\approx 5^0)$, while
varying $\d$ under the condition $0.025 \le R_\n\le 0.049$,  we find that all successful
models turn out to be  tightly constrained in order to have a quasi
degenerate spectrum, ($m_1\sim m_2$), for $m_1$ and $m_2$ and no strong hierarchy
between $m_1\sim m_2$ and $m_3$ can occur. This can be considered as a general
prediction for these class of models.

\section*{Acknowledgement}
One of the authors, E. I. Lashin would like to thank both of A. Smirnov and S. Petcov
for useful discussions. Part of this work was done within the associate scheme of ICTP.
%
\renewcommand{\baselinestretch}{1}

\end{document}